\documentclass[manuscript]{aastex}

\shorttitle{Stellar Astrophysics with a dFTS. I.}
\shortauthors{Behr et al.}

\begin{document}

\title{Stellar Astrophysics with a \\
Dispersed Fourier Transform Spectrograph. \\
I. Instrument Description and Orbits of \\
Single-lined Spectroscopic Binaries}

\author{Bradford B. Behr\altaffilmark{1,2},
Arsen R. Hajian\altaffilmark{1},
Andrew T. Cenko\altaffilmark{1}, 
Marc Murison\altaffilmark{3},
Robert S. McMillan\altaffilmark{4},
Robert Hindsley\altaffilmark{5}, and
Jeff Meade\altaffilmark{1}}

\altaffiltext{1}{Department of Systems Design Engineering, University of Waterloo, Waterloo ON N2L 3G1, Canada.}
\altaffiltext{2}{Eureka Scientific, 2452 Delmer Street Suite 100, Oakland, CA 94602-3017.}
\altaffiltext{3}{US Naval Observatory, Flagstaff Station, 10391 W.\ Naval Observatory Rd., Flagstaff, AZ, 86001.}
\altaffiltext{4}{Lunar and Planetary Laboratory, University of Arizona, Tucson, Arizona 85721.}
\altaffiltext{5}{Remote Sensing Division, Naval Research Laboratory, Code 7215, Washington, DC, 20375.}


\begin{abstract}

We have designed and constructed a second-generation version of the Dispersed Fourier Transform Spectrograph, or dFTS. This instrument combines a spectral interferometer with a dispersive spectrograph to provide high-accuracy, high-resolution optical spectra of stellar targets. The new version, dFTS2, is based upon the design of our prototype, with several modifications to improve the system throughput and performance. We deployed dFTS2 to the Steward Observatory 2.3-meter Bok Telescope from June 2007 to June 2008, and undertook an observing program on spectroscopic binary stars, with the goal of constraining the velocity amplitude $K$ of the binary orbits with 0.1\% accuracy, a significant improvement over most of the orbits reported in the literature. We present results for radial velocity reference stars and orbit solutions for single-lined spectroscopic binaries.

\end{abstract}



\section{Introduction}

For high-resolution ($R = \lambda/\Delta\lambda \simeq$~50,000) spectroscopy of stellar targets, most astronomers use cross-dispersed echelle spectrographs.  When equipped with an iodine absorption cell or thorium-argon calibration source, echelle spectrographs can measure the radial velocity (RV) of a star with a precision of $\sim 1$~m/s \citep{bou09, how09}. This approach to stellar velocimetry has proven highly successful in the hunt for exoplanets and other low-mass, low-luminosity companions orbiting bright stars.

However, the application of these ``PRV'' (precision radial velocity) techniques to binary stars has been more limited. Models of stellar structure and evolution require accurate measurement of stellar masses for validation and refinement, and binary star systems are among the best celestial laboratories for making such measurements. Given the period $P$, radial velocity amplitudes $K_1$ and $K_2$ of the two components, eccentricity $e$, and inclination angle $i$ of a double-lined spectroscopic binary (SB2) system, the masses of both stars can easily be derived. If the angular size of the semimajor axes of the orbit can also be measured, using spatial or speckle interferometry, then the distance to the system can be determined as well.

When an SB2 target is observed with an iodine-absorption echelle, it becomes algorithmically challenging to disentangle the three superposed spectra (two different stars plus the iodine absorption lines). \citet{kon05, kon09} has made notable progress towards solving this challenge, and in conjunction with astrometric data from spatial interferometers, his technique has yielded some very precise binary star masses \citep{mut06}, but this approach has not yet been widely adopted.

Thorium-argon emission line lamps provide another means of precisely determining the wavelength scale for echelle spectra, and several groups have achieved excellent RV results on spectroscopic binaries using Th-Ar calibration. \citet{tom06} report RMS velocity residuals of 0.11~km/s on two SB2 systems, a particularly notable accomplishment given that their spectrograph, the Sandiford Echelle \citep{mcc93}, is mounted at the Cassegrain focus of the telescope, and is thus subject to a changing gravity vector. \citet{fek07} achieve a similar level of RV precision with a fiber-fed, bench-mounted echelle. The HERCULES instrument \citep{hea02}, a fiber-fed echelle mounted inside a vacuum chamber for greater RV stability, has achieved impressive RMS velocity residuals of 14 to 56~m/s on binary targets \citep{ram04, sku04, ram08}.

Our research group has been pursuing an alternative approach to stellar velocimetry, developing a new instrument which provides meter-per-second RV accuracy without a superposed reference spectrum or vacuum enclosure. This instrument concept is known as the Dispersed Fourier Transform Spectrograph, or dFTS. Our prototype device, dFTS1, is described in detail by \citet{haj07} (henceforth, ``Paper 1''). Installed at the Clay Center Observatory 0.6-meter telescope, dFTS1 demonstrated the viability of the concept, with precision RV measurements of spectroscopic binary systems and exoplanet host stars.

To develop the dFTS concept further and undertake a comprehensive science program, we designed and constructed a new device, called dFTS2, and deployed it at the Steward Observatory 2.3-meter Bok Telescope on Kitt Peak. In this paper, we describe the dFTS2 instrument and present results from our observing campaign, focusing on RV reference stars and single-lined spectroscopic binary stars (SB1s). Single-lined binaries are not as scientifically interesting as SB2s, because absolute stellar masses cannot be derived, but the SB1s still serve as useful tests of our spectrograph performance. Orbital elements of double-lined systems will be discussed in a subsequent paper.


\section{Overview of dFTS concept}

The dFTS concept and its underlying mathematical theory are treated in detail in Paper 1, but in the following section we review the some of the key aspects of the design.

A traditional Fourier Transform Spectrograph (FTS) consists of a Michelson interferometer with one retroreflector mounted on a translation stage (Figure~1, left). An input source is collimated and then divided into two equal parts by a beamsplitter cube (BSCA). Each beam travels down one ``arm'' of the interferometer, hits a corner-cube retroreflector (RR1 or RR2), and returns to a second beamsplitter (BSCB), where the two beams combine interferometrically. Depending on the wavelength of the light and the optical path difference (OPD) or ``delay'' between the two arms, the outputs from BSCB will exhibit constructive or destructive interference. (In the figure, only one output is shown; a complementary output exits the top face of BSCB.)

As RR2 is moved on its translation stage, the delay changes, and the intensity of the interferometer output changes as well. Normally, the retroreflector stage is scanned or stepped through a series of different positions, and the output intensity at each delay is measured using a photosensitive detector (D1). The result is an interferogram (Figure~1, bottom right), which is the Fourier transform of the spectral energy distribution of the input source (Figure 1, top right). By performing a Fourier inversion of the interferogram, the source spectrum can be recovered. For a higher-resolution spectrum, the interferometer merely needs to scan over a wider range of delay, and the wavelength scale of the spectrum can be derived very accurately if the delay positions are measured with a metrology system.

However, if the input source is broadband, as depicted in Figure 1, then the amplitude of a FTS's interferometric ``fringes'' (the oscillations of intensity as a function of delay) drops off quickly as we move away from the central delay position (OPD~$= 0$), because all the different wavelengths quickly decorrelate. Small-amplitude fringes are still present, but they become very difficult to detect given measurement noise. Fine details of the input spectrum, which are encoded in the high-delay fringes, are thus lost.

To boost the fringe contrast, a narrowband filter (NBF) can be added to a FTS, thereby restricting the input bandpass (Figure 2). The interferometer output remains coherent over a wider range of delays, with stronger fringes at large delay, so that details of the spectrum can be measured. However, all of the light outside of the filter bandpass is discarded. For some astronomical applications, where only a narrow spectral region is of interest, this is acceptable, but for broadband spectroscopy, the overall system efficiency is unacceptably low. Traditional FTS devices have therefore seen only limited use in stellar astronomy.

\begin{figure}[t]
\centering
\includegraphics[scale=0.7]{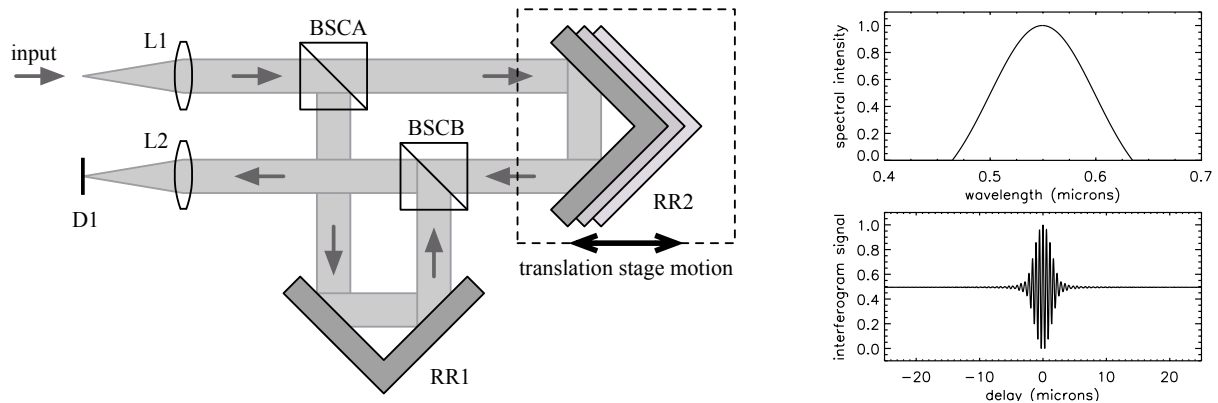}
\caption{Schematic of a conventional FTS using an offset Michelson interferometer configuration (left), with a broadband input spectrum (top right) and resulting interferogram (bottom right).}
\rule{150pt}{0.5pt}
\end{figure}

\begin{figure}[t]
\centering
\includegraphics[scale=0.7]{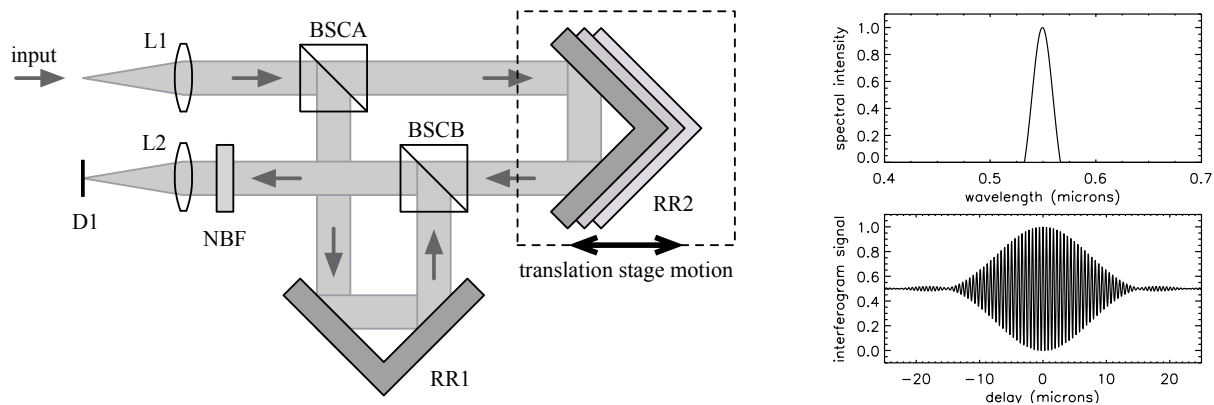}
\caption{Schematic of a conventional FTS with a narrowband filter (left). The narrowed spectral bandpass (top right) results in a broader central fringe packet in the interferogram (bottom right), such that the fringes at large delay are stronger than in Figure 1.}
\rule{150pt}{0.5pt}
\end{figure}

\begin{figure}[t]
\centering
\includegraphics[scale=0.7]{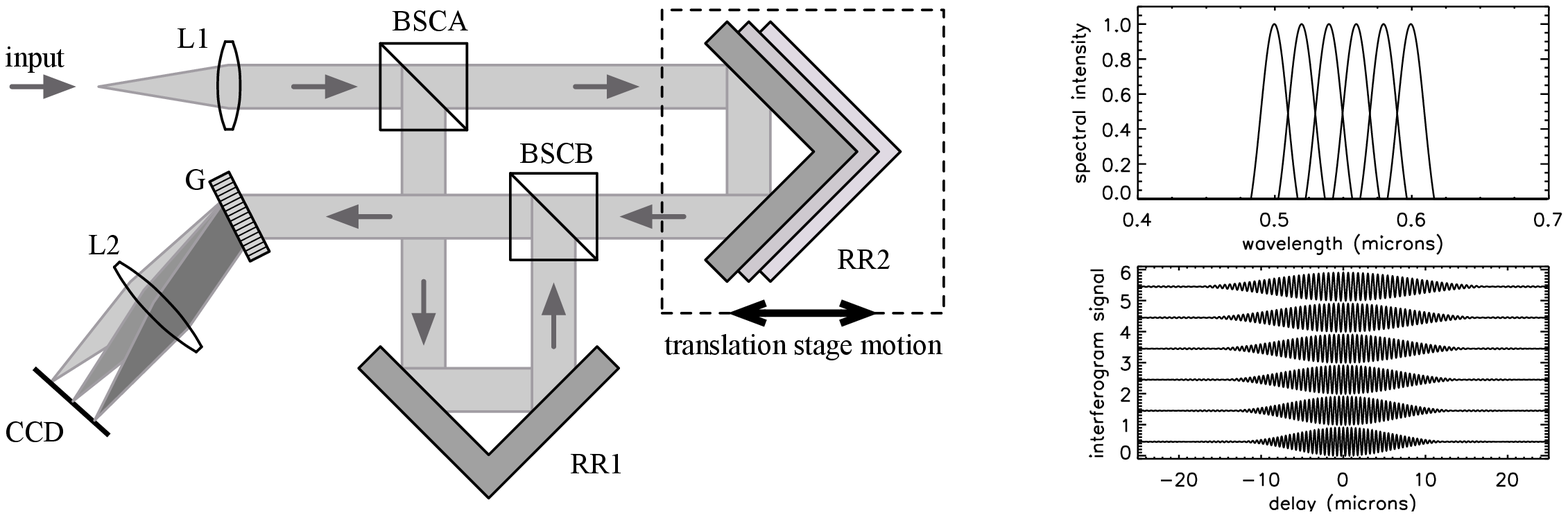}
\caption{Schematic of a dispersed FTS (left). The Michelson interferometer is followed by a grating spectrograph, which acts as a multiplexer. Each pixel along the CCDÕs dispersion axis acts as the detector for a separate spectral channel (top right). Because the individual channels are narrowband, their interferograms (bottom right) have wide central fringe packets and stronger fringes at large delay than in the broadband case.}
\rule{150pt}{0.5pt}
\end{figure}

The dFTS concept addresses this limitation of the traditional FTS design. Figure 3 depicts a dFTS configuration in schematic form. The narrowband filter has been replaced by a dispersive grating (G), and the dispersed spectrum is focused onto an array detector (CCD). The broadband output beam from the interferometer is thus divided into many separate spectral channels (Figure 3, top right), each of them covering just a small subsection of the initial bandpass. The figure shows only six channels, but in an actual dFTS instrument, there would be thousands of channels, one for each CCD pixel along the dispersion dimension. Because the channels are narrowband, their fringe patterns (Figure 3, bottom right) span a wide range of delays, and we can derive the fine details of the spectrum from the high-OPD fringes without sacrificing spectral coverage. In essence, the grating acts as a multiplexer, converting a single broadband FTS into several thousand narrowband FTSs, all operating in parallel and sharing the same interferometer optics. Each channel interferogram transforms into a narrowband spectrum. Merging the narrowband spectra produces a high-resolution broadband spectrum.

An additional advantage of narrowband output channels is that the interferogram can be sampled much more coarsely. The Nyquist theorem requires that, in order to avoid aliasing, the fringes must be sampled with a delay step size $\Delta x$ which is less than or equal to $1/(2\Delta s)$, where $\Delta s$ is the bandwidth in wavenumbers. With narrowband channels, therefore, we can cover a wide range of delays (and thus derive a high-resolution spectrum) with relatively few delay positions. This ``sparse-sampling'' strategy is treated in more detail in the Appendices of Paper 1.   

Two additional strengths of the dFTS concept should be noted. First, the instrumental broadening function (or line spread function) can be determined {\it a priori} from the delay sampling function, so that the true shape of the stellar absorption line profiles can be derived in a straightforward fashion. Second, because the high-resolution information in the source spectrum is extracted by the interferometer, the dispersive part of the dFTS does not have to be as powerful as in an echelle ({\it e.g.} $R = 5$,000 instead of $R = 50$,000), so the collimated beam diameter at the disperser can be considerably smaller. This issue is particularly relevant for instruments on 8--10 meter telescopes, where the instrument optics can become exceedingly large and expensive. Smaller-diameter optics are also advantageous for instruments in spacecraft, airborne platforms, and field-portable sensor packages.


\section{Instrument design and deployment}

\subsection{Design considerations}

The dFTS1 prototype proved successful for validating the instrument concept and demonstrating its capabilities for accurate measurement of stellar RVs, but in the course of commissioning and operating dFTS1 we identified several potential improvements to the optical design and implementation. The design of dFTS2 was thus motivated by five main drivers:
\begin{enumerate}
\item Higher photon throughput.
\item Higher spectral resolution $R_{\rm G}$ of the dispersive backend, for narrower channel bandpasses and higher fringe contrast over the delay range.
\item Good coupling to 2-meter telescope optics.
\item Smaller physical size, for transportation to observing facilities.
\item Utilization of optics and optomechanical components that were already in hand, or could be acquired inexpensively, to fit within our budget constraints.
\end{enumerate}

\subsection{Final instrument design}

The final configuration of dFTS2 is shown in Figure 4. The interior of the instrument's thermal enclosure measures approximately 100 cm $\times$ 70 cm, with a height of 30 cm. The interferometer and laser metrology system are mounted on a larger optical breadboard (on the left side of the photograph in Figure 4), while the dispersive backend is placed on a smaller separate breadboard (on the right side), so that vibrations from the mechancal shutter and CCD fans would not affect the interferometer.

Starlight from the telescope enters the instrument via a multimode fiber (MMF). The beam exiting the fiber is collimated by lens L1 and passes through an iris with 20~mm aperture. It then encounters BSC1, the first of four polarizing beamsplitter cubes. BSC1 is mounted at a $45^\circ$ angle from the plane of the breadboard, so the transmitted light is diagonally (linearly) polarized. At BSC2 (which corresponds to BSCA in the prior figures), the vertically-polarized component of the beam is reflected towards stationary retroreflector RR1, while the horizontally-polarized part transmits and goes to moving retroreflector RR2. For translating RR2, we use an Aerotech ANT-50L linear motor stage, instead of the Parker Daedal MX-80 stage in dFTS1 (as described in Paper~1). The ANT-50L provides better stability while holding at a fixed delay position; the RMS jitter in optical delay position is typically less than 20~nm.

\begin{figure}[p]
\centering
\includegraphics[scale=0.145]{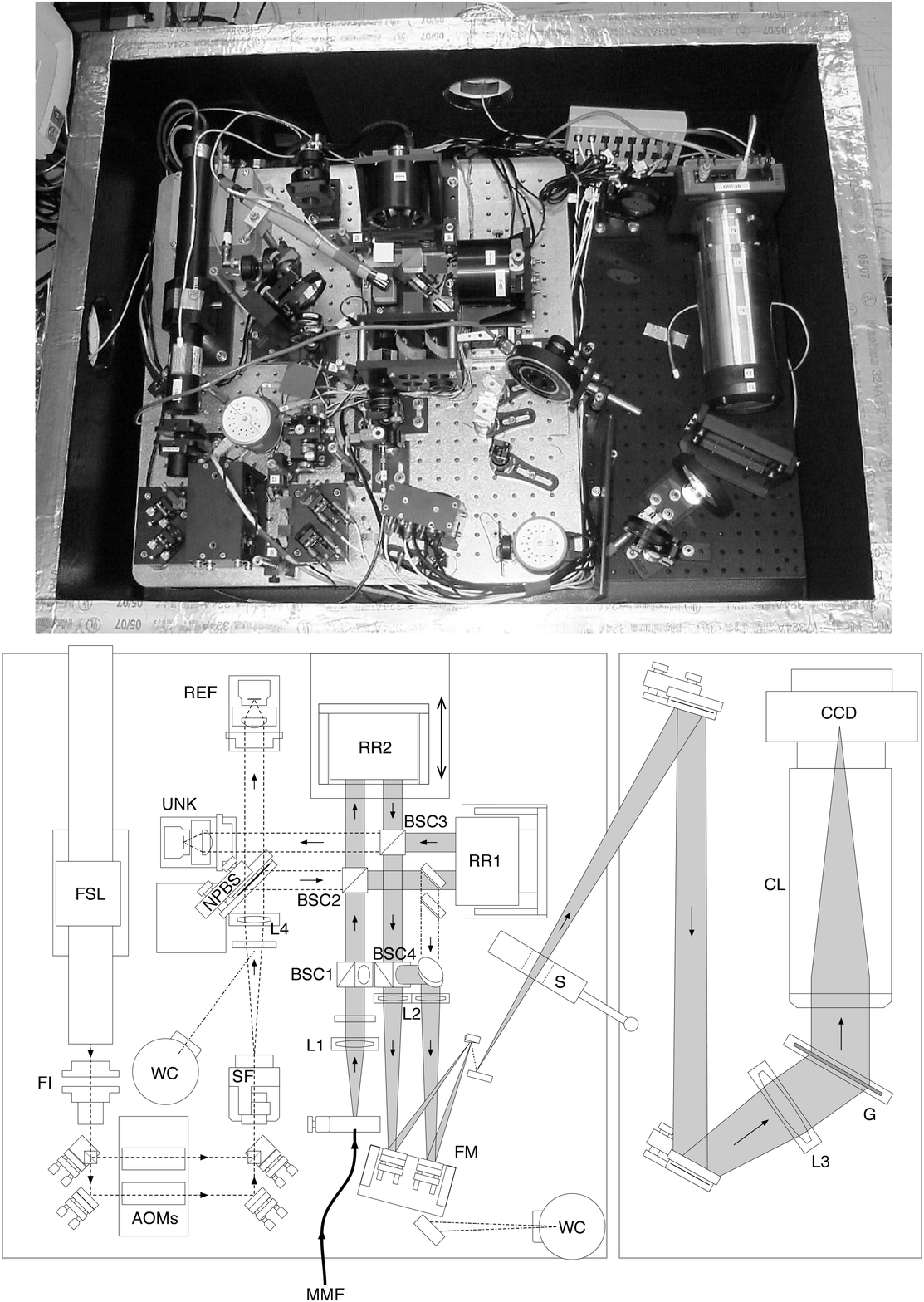}
\caption{Photograph and layout schematic of the final dFTS2 design. Grey sections show the starlight beams, dashed lines delineate the laser metrology beams, and dot-dash lines indicate beam paths for optical alignment. See the text for explanation of the component codes.}
\end{figure}

After retroreflection, the beams meet at BSC3 (equivalent to BSCB), where the vertically-polarized beam is again reflected and the horizontally-polarized beam is transmitted, such that the beams are once more coincident and parallel. Because of their orthogonal polarizations, however, they cannot combine interferometrically until they encounter BSC4, which is mounted at $45^\circ$ like BSC1 and thus ``mixes'' the two polarizations together. The two outputs from BSC4 (one transmitted, one reflected) are each focused by a lens L2, are steered by a series of fold mirrors (FM), and are passed through the shutter S en route to the dispersive backend of the instrument. The light that is reflected from BSC1 (the ``C-beam'') is focused by a third L2 lens, and also enters the backend, where it provides a non-interferometric flux reference.

We had originally placed a set of optical fiber spot-to-line bundles just prior to the shutter, in order to ``slice'' the output beams from the interferometer, and thus achieve higher grating resolution $R_{\rm G}$ in the dispersive backend. We used this fiber arrangement for the first observing run but found that it was very inefficient, so we replaced it with an arrangement of small mirrors before the second observing run. 

The three diverging beams are collimated by achromat L3 and then dispersed by grating G, a Kaiser Optical Systems Inc. volume-phase holographic grating with 1,800 lines/mm. For the camera lens (CL), one of us (Murison) designed a custom f/3.6 camera lens with 180 mm focal length, using only off-the-shelf singlets and achromats. When installed prior to the second observing run, this custom lens provided a vast improvement in photon efficiency over the Nikon SLR lens that we had been using previously. The camera lens is bolted to an Apogee U-1107 CCD, with a $2048 \times 122$ array of $12\mu{\rm m} \times 12\mu{\rm m}$ pixels. The total instrument bandpass spans 470~nm to 540~nm.

As with dFTS1, a laser metrology system continuously monitors the optical delay during a data acquisition scan, so that each delay position is precisely and unambiguously known. A Melles-Griot 05-STP-901 frequency-stabilized laser (FSL) acts as the wavelength reference. The laser beam passes through a Faraday isolator (to suppress backscatter) and is then divided into horizontally- and vertically-polarized components by a small polarizing BSC. These beams are frequency-shifted by two acousto-optic modulators (AOMs), which are driven at two slightly different frequencies, creating an 11 kHz beat frequency between the horizontal and vertical polarizations. The two polarizations are recombined by a second small BSC, and then sent through a spatial filter (SF) to produce a flat and uniform wavefront. Lens L4 collimates the metrology beam to a diameter of 20~mm, matching the starlight beam size. A non-polarizing 50/50 beamsplitter (NPBS) sends half of the metrology beam into the interferometer, while the other half goes to a ``reference'' metrology detector (REF), where the horizontal and vertical components of the beam are mixed together with a linear polarizer, and the resulting 11 kHz sinusoidal interference signal is detected by a photodiode. The metrology beam enters the interferometer at BSC2, following exactly the same path as the starlight beam until BSC3, where it exits and is captured by the ``unknown'' metrology detector unit (UNK), which also detects the 11 kHz sinusoidal signal. A change in the interferometer delay manifests as a phase shift between the REF and UNK signals. This phase difference is measured and recorded, and then converted into an absolute delay position in postprocessing.

The dFTS2 instrument is housed inside a thermal enclosure, with a temperature stabilization system that keeps the internal air temperature constant to $\pm 0.05^\circ$C. A Vaisala PTU-200 atmospheric monitoring unit measures the temperature, pressure, and humidity of the airmass in the immediate vicinity of the interferometer, so that we can correct our laser metrology data to account for changes in the index of refraction, which would otherwise induce shifts of $\sim 20$~m/s in the RV zero point of the instrument. Key optical components are placed on tip-tilt mounts with New Focus ``Picomotor'' actuators, for remote adjustment, and critical beam alignment positions are viewed remotely with D-Link DCS-900 webcams (WC).

\subsection{Telescope guider box and fiber feed}

Starlight reaches the dFTS2 instrument via a multimode optical fiber from a customized ``guider box'' that bolts to the telescope's Cassegrain focus position. The f/9 converging beam from the telescope's secondary mirror reflects from a SBIG AO-7 tip-tilt mirror and then passes through a small achromat lens, which speeds up the beam to f/2.5. Just before encountering the input tip of the fiber, the beam passes through an uncoated pellicle (92\%\ transmission). The $50\mu$m core of the fiber subtends 1.8 arcseconds on the sky. The light that does not enter the fiber is reflected by the polished face of the fiber ferrule, and 8\%\ of it bounces off the pellicle and is reimaged on to an Astrovid StellaCam III video camera, which provides an image for guiding. This video signal is digitized and monitored by a custom software package, which adjusts the AO-7 tip-tilt mirror to keep the star image centered on the fiber core.

Light from calibration sources (such as an incandescent lamp or a hollow-cathode emission-line lamp) is sent to the guider box through a secondary optical fiber. The output from this fiber is reimaged onto the primary fiber input via a fold mirror which is inserted into the telescope beam path, so that the calibration source reaches the main fiber from the same direction and at the same f/ratio as starlight. This feature of dFTS2 ensures that the instrument illumination pattern is the same for calibration and science exposures.

\subsection{Deployment and operation}

The dFTS2 hardware was assembled at the U.S. Naval Observatory in Washington, D.C., and after a brief testing period was shipped to Steward Observatory's 2.3-meter Bok telescope on Kitt Peak in Arizona. For the initial commissioning run, we installed the instrument and support electronics in a storage room one floor below the telescope, but this room was not air-conditioned, and the instrument was prone to overheating, so for subsequent runs we moved dFTS2 to the control room next to the telescope. The thermal environment was more stable at this location, although the temperature stabilization system still had to contend with fluctuations of $\sim 4^\circ$C in the ambient air temperature as the room's A/C or heating system cycled.

We had eight observing runs on the Bok telescope with dFTS2, on the following dates (civil) : 27 Jun -- 04 Jul, 2007; 30 Sep -- 03 Oct, 2007; 27 Oct -- 31 Oct, 2007; 22 Jan -- 27 Jan, 2008; 22 Mar -- 24 Mar, 2008; 20 Apr -- 23 Apr, 2008; 16 May -- 20 May, 2008; and 17 Jun -- 21 Jun, 2008.

Interferometric scans on stellar targets consisted of 500 delay step positions, with a 0.25 to 4.0 second exposure at each position, depending on the star's brightness and the atmospheric seeing and opacity.


\section{Instrument performance}

\subsection{Photon throughput}

To test the real-world photon efficiency of the instrument, we measured the photoelectron detection rate on our CCD while observing photometrically stable stars under good seeing conditions, and compared those numbers to the expected photon flux as determined from the stars' published V magnitudes. We estimate the total system efficiency (including atmospheric opacity, telescope mirror reflectivity, fiber feed transmission, instrument mirror reflectivity, and CCD quantum efficiency) to be 4.3\%. This figure represents a significant ($6\times$) improvement over the 0.7\% total efficiency we reported for dFTS1. The enhanced throughput performance of dFTS2 is due primarily to better coupling into the fiber in the telescope guider box, a simpler instrument layout with fewer fold mirror reflections, and the custom camera lens. In conjunction with the larger collecting area of the Bok telescope, these upgrades have permitted us to achieve much better RV precision on fainter stars than with dFTS1.

However, further improvements in photon throughput are certainly possible. The 50$\mu$m core diameter of the optical fiber was not well-matched to the typical seeing disk size at the Bok telescope; with a 100$\mu$m fiber and suitable modifications to the downstream optics, we might have achieved a 50--80\% increase in the measured stellar flux. The optics in the dFTS2 instrument box also accumulated some visible dust over the course of our observing campaign, despite our efforts to keep the system enclosed as much as possible during assembly and disassembly. With future dFTSes permanently installed as facility instruments, we anticipate being able to reach photon throughput efficiencies of 10 to 12\%, similar to many echelle spectrographs.

\subsection{Dispersive backend resolution}

During each night of observing, we collected an interferometric scan of a incandescent white light source, which let us measure the central wavelength and the bandwidth of each spectral channel on the CCD. These channel bandpass data are subsequently used in reconstructing the broadband high-resolution spectra  of our stellar targets and measuring their RVs. We can also use these white-light scans to evaluate the performance of the dispersive backend system. We measure a spectral resolution $R_{\rm G} > 4$,700 over the entire spectral range of dFTS2, with a mean $R_{\rm G}$ of approximately 4,950 and a peak of 5,150. This resolution is nearly three times better than the $R_{\rm G} \simeq 1$,700 of dFTS1, which results in broader fringe packets, greater fringe visibility, and thus better SNR of the final high-resolution spectra.

\subsection{Intrinsic RV stability}

In addition to the white-light scans, we also acquired at least six scans of a hollow-cathode thorium-neon emission line lamp per night. (The only available thorium-argon lamp was not bright enough for reliable measurements.) By measuring the interferometric fringe patterns of the $\sim 60$ brightest emission lines, we were able to evaluate the intrinsic RV stability of the instrument. These calibration measurements also defined the RV zero point for each night of an observing run; when the instrument was reassembled at the beginning of each run, or realigned during a run, the alignment between the starlight and metrology beams could have shifted slightly, creating a systematic offset in derived RV.

Figure~5 shows the measured RVs of the thorium-neon source for 224 scans over the course of our observing program, from September 2007 to June 2008. On some nights, such as 2008-01-24 and 2008-04-22, the instrument was quite stable, with a $\sigma({\rm RV})$ of 1.5 m/s or better. On other nights, however, we see much greater variability, with $\sigma({\rm RV})$ of approximately 10~m/s and peak-to-peak fluctuations as large as 20 m/s. The ``unstable'' nights seem to be due to temperature fluctuations of $\sim 0.2^\circ$C of the optomechanical structure of the instrument, induced by $\sim 5^\circ$C changes of the ambient air temperature in the room housing the dFTS2 hardware. Even though the air temperature inside the dFTS2 enclosure is controlled to $\pm 0.05^\circ$C, the optical breadboard and optics mounts can get warmer or cooler and change their shape enough to induce a measurable RV shift. To account for this error source, we apply an error of 10~m/s (added in quadrature) to the radial velocity error estimates for all our stellar observations, as well as adjusting the RV zero point on a night-by-night basis (see below). Future dFTS instruments will need to be installed in more thermally stable environments to achieve their full RV measurement potential.

\begin{figure}[t]
\centering
\includegraphics[scale=0.7]{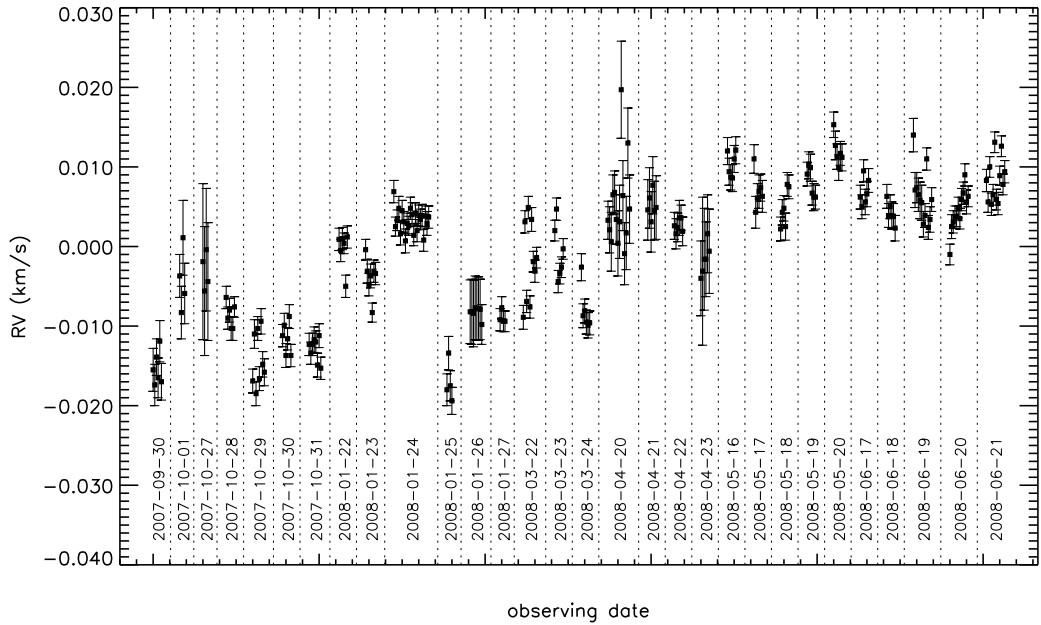}
\caption{Night-to-night radial velocity zero point of dFTS2, measured using a thorium-neon emission line lamp.}
\rule{150pt}{0.5pt}
\end{figure}


\section{Data reduction and derivation of radial velocities}

Reduction of the interferogram data from dFTS2 followed the same basic pathway as for dFTS1 (see section 3.4 of Paper 1), with the addition of an intermediate step to correct for a significant nonlinearity in the response of the U-1107 CCD. Photometric corrections were particularly important for these data, as seeing fluctuations and passing thin clouds would often change the stellar flux by a factor of 2 or 3 from one exposure to the next. By using the unfringed ``C-track,'' which was picked off from the main starlight beam before it entered the interferometer, we were able to normalize the flux levels over the course of a scan, so that only true interferometric fringes remained.

To determine a topocentric radial velocity for each observation, we compared the normalized interferograms to synthetic interferograms derived from a Doppler-shifted template spectrum. We sampled the synthetic interferograms at the same delays as each observation scan, and we scanned through a range of RVs until we found a smooth global minimum in the $\chi^2$ difference between the data and the model. This technique incorporated the optimal weighting of the data and provided an estimate of the internal statistical error in RV for each observation.

We employed two different types of template spectra for these RV fits. For the first template, we used the FROID algorithm, described in detail in Paper 1, to turn the parallel narrowband interferograms into a single high-resolution ($R =$~50,000) broadband spectrum for each observation. All derived spectra for a given star were then shifted to zero Doppler velocity and coadded to make a template specific to that star. The second template type was generated using the SPECTRUM spectral synthesis package \citep{gra94} (see also \url{http://www.phys.appstate.edu/spectrum/spectrum.html}) and ATLAS9 model atmosphere grids \citep{kur93}. We found that spectra generated using the default atomic line list resulted in sub-optimal RV solutions, because the depths of the synthetic absorption lines were not perfectly matched to the actual spectra. By adjusting the transition strength $\log g\!f$ of each line individually (sometimes increasing, sometimes decreasing), we were able to create synthetic templates which performed as well as the coadded templates. All of the RV results presented below were derived using synthetic spectral templates instead of coadded templates, but the RV results were very similar in all cases.

As a final step, we converted the topocentric velocities into barycentric velocities using the IRAF tool BCVCORR. We also applied an additional velocity correction factor based on the mean thorium-neon velocity offset for each night of observing, to account for RV shifts due to instrument realignment. We did not make any adjustment for the varying light travel time within the binary systems, since the resulting change in RV is small compared to the RV measurement errors.


\section{RV measurements of reference stars}

To verify that dFTS2 was stable, we observed three astronomical targets which were expected to maintain a constant barycentric radial velocity over time.

\subsection{$\beta$ Ophiuchi}

The star $\beta$ Ophiuchi (HR~6603, HD~161096, HIP~86742) is a K2III giant with an apparent $V$ magnitude of 2.77 and no known binary companion. We made 30 observations of $\beta$~Oph between March and June 2008, and the heliocentric RV data are plotted in Figure~6. The RV points exhibit an RMS scatter of 13.6 m/s around the mean. The RV error is dominated by the temperature-induced instrument instability mentioned in Section 4.3. Without this source of systematic error, we estimate that we could have achieved a mean RV accuracy of 2.1~m/s per observation on $\beta$~Oph, based upon $\chi^2$ analysis of photon statistics in the interferogram data. 

\begin{figure}[ht]
\centering
\includegraphics[scale=0.7]{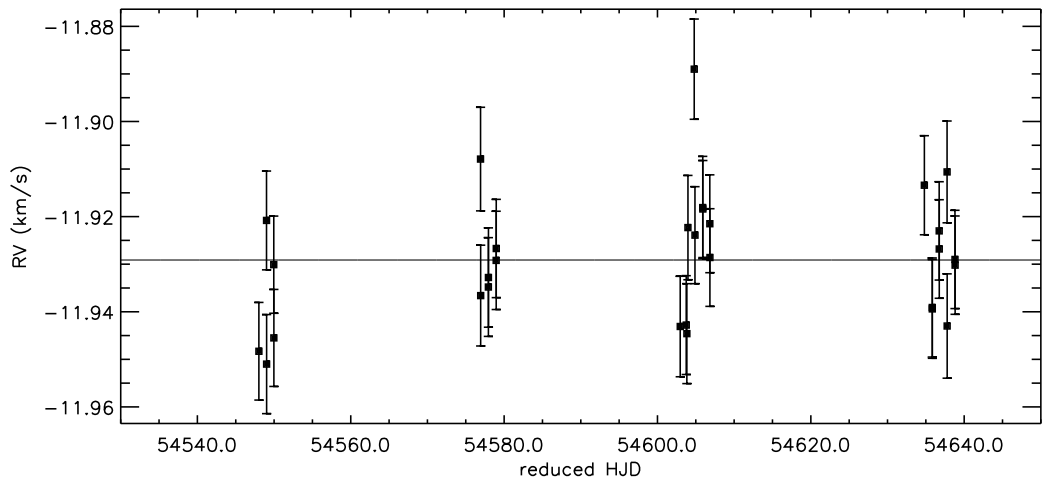}
\caption{Radial velocity measurements of the K2III star $\beta$ Ophiuchi, with an RMS scatter of 13.6~m/s.}
\end{figure}

\subsection{$\beta$ Canum Venaticorum}
The G0V star $\beta$ Canum Venaticorum (HR~4785, HD~109358, HIP~61317) was our second RV reference star. At $V = 4.26$, it tested the RV performance of our instrument at lower flux levels. Figure~7 depicts our RV results. We find an RMS scatter of the barycentric RV points of 13.8~m/s, similar to that of $\beta$~Oph because the error budget is dominated by the same systematic error sources. If we were limited solely by photon statistics and CCD read noise, we estimate that dFTS2 could achieve a per-observation RV accuracy of 5.1~m/s on this star. 

\begin{figure}[ht]
\centering
\includegraphics[scale=0.7]{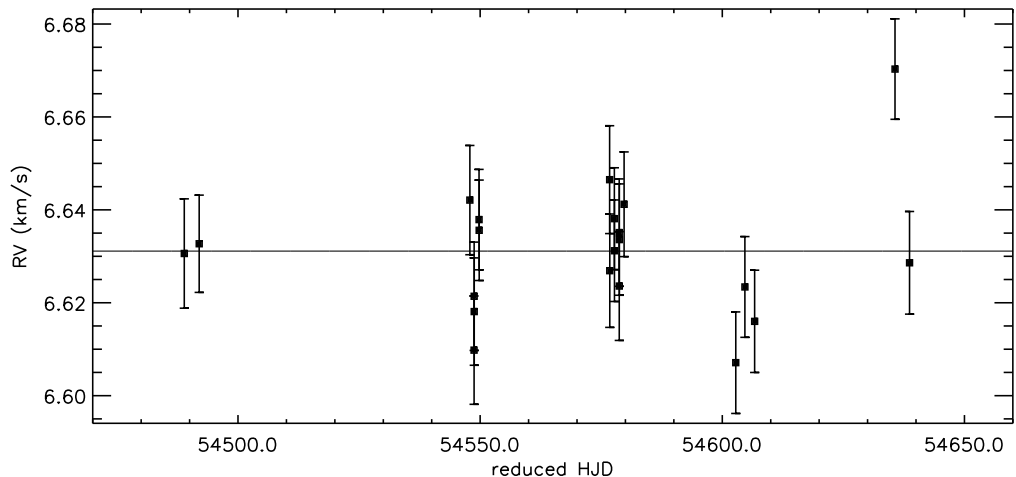}
\caption{Radial velocity measurements of the G0V star $\beta$ Canum Venaticorum, with an RMS scatter of 13.8~m/s.}
\end{figure}

\subsection{The Moon}
As our third RV test target, we observed sunlight reflected from the bright lunar crater M\"osting~A, which provided a flux level similar to a $V = 2.8$ star. We used lunar ephemerides calculated with the MICA package \citep{oli05} to shift the measured velocities to a common reference frame. The resulting time series of velocities, shown in Figure~8, has an RMS scatter of 9.7~m/s. The mean internal error bar from photon statistics is 2.5 m/s for these observations.

\begin{figure}[ht]
\centering
\includegraphics[scale=0.7]{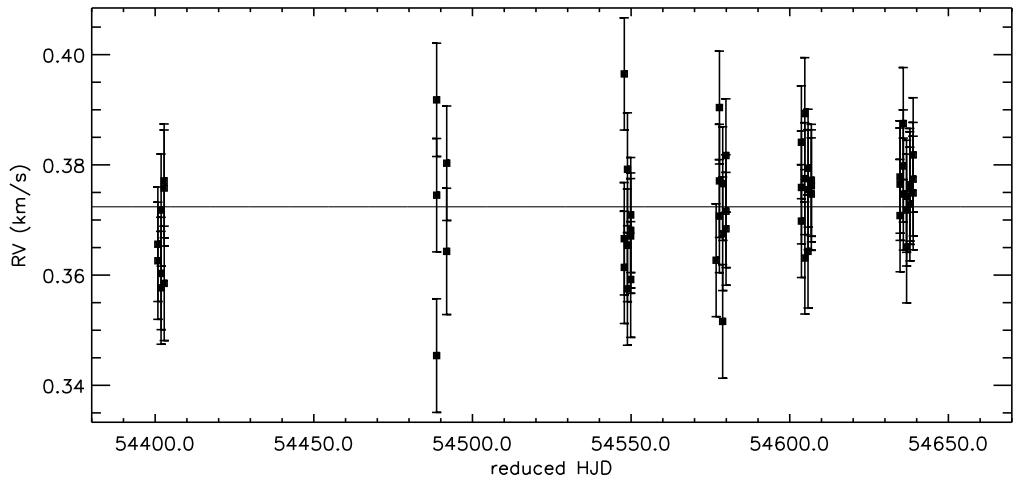}
\caption{Radial velocity measurements of the G2V solar spectrum reflected from the lunar crater M\"osting~A. The RMS scatter of the velocity points is 9.7 m/s.}
\end{figure}


\section{RV measurements of SB1 targets}

During our time at the Bok telescope, we observed a variety of single-lined (SB1) and double-lined (SB2) spectroscopic binary systems. The results for the SB1s are described below, while the orbital fits for the SB2s will be reported in a following publication. We had hoped to collect $\sim 40$ RV points for each binary, well-distributed over the orbital phase, in order to derive the binary parameters with higher accuracy than ever before. Unfortunately, bureaucratic considerations led to the decommissioning of the dFTS2 instrument in August 2008, and thus the end of our observing program. As a result, none of our orbital RV curves are well sampled, and for systems with $P > 100$~days we do not even cover a full orbit. The orbital parameters that we derive from our RV measurements should therefore be considered as preliminary results, which serve primarily to illustrate the potential usefulness of dFTS instrumentation for the accurate determination of binary star orbits.

For deriving orbital parameters from our RV data, we used the IDL routines {\tt CURVEFIT}, a nonlinear least-squares fitting algorithm included with the IDL package, and {\tt HELIO\_RV}, which computes binary star orbit velocities \citep{lan93}. In all cases, we adopted the orbital period $P$ from previously-published analyses, because we had too few data points to derive an accurate value. 

\subsection{19 Draconis}

The 19~Draconis system (HR~6315, HD~153597, HIP~82860) consists of a F6V star with $V = 4.89$ co-orbiting with an unseen companion. With only nine RV observations, we could not constrain the orbital period, so we adopted $P = 52.1089$~days from \citet{abt76} but let all other orbital parameters vary in the fitting algorithm. Figure~9 shows the RV points and the resulting orbit, with parameters $T_{\rm peri} = 54522.258 \pm 0.121$~days, $V_0 = -20.5299 \pm 0.0649$~km/s, $K = 17.1590 \pm 0.0344$~km/s, $e = 0.2218 \pm 0.0020$, and $\omega = 339.1^\circ \pm 0.6^\circ$. The uncertainty on the value of $K$ yields a fractional uncertainty $\Delta K/K$ of 0.20\%.

The RV residuals show an RMS scatter of 27.5 m/s, while the mean formal error bar was 21.3 m/s, with $\chi^2 = 12.19$ for $\nu = 4$. We used the formal error bars in calculating the parameter uncertainties cited above. If these error bars are scaled such that the mean error bar is equal to the RMS scatter, then the uncertainty on $K$ increases to 0.0444~km/s, or $\Delta K/K = 0.26$\%. With relatively few data points, it is difficult to tell whether this discrepancy between the formal RV errors and the RV residual RMS is just a statistical aberration, or is indicative of an additional RV error source, such as stellar pulsation or the spectral contribution of the faint secondary. As pointed out by our anonymous referee, contamination from the secondary spectrum can add noise to RV measurements of the primary. Using the Hipparcos parallax \citep{per97} for 19~Dra, we calculate an absolute magnitude of $M_V = 4.00$, or $L_1 \simeq 2.15 L_\odot$, assuming that the primary dominates the total system flux. Based on the F6V spectral type, we estimate $M_1 \simeq 1.26 M_\odot$. Using the mass function for the system and $i = 56.1^\circ$ from \citet{jan05}, we can then estimate $M_2 \simeq 0.54 M_\odot$, or $L_2 \simeq 0.12 L_\odot$ for a main sequence star. The secondary might therefore account for 6\% of the total flux in V-band, which could be enough to influence the measured velocities for the primary, and might also be detectable. Future analysis work will attempt to detect the secondary spectrum, or place more stringent limits on its brightness. 

\begin{figure}[ht]
\centering
\includegraphics[scale=0.7]{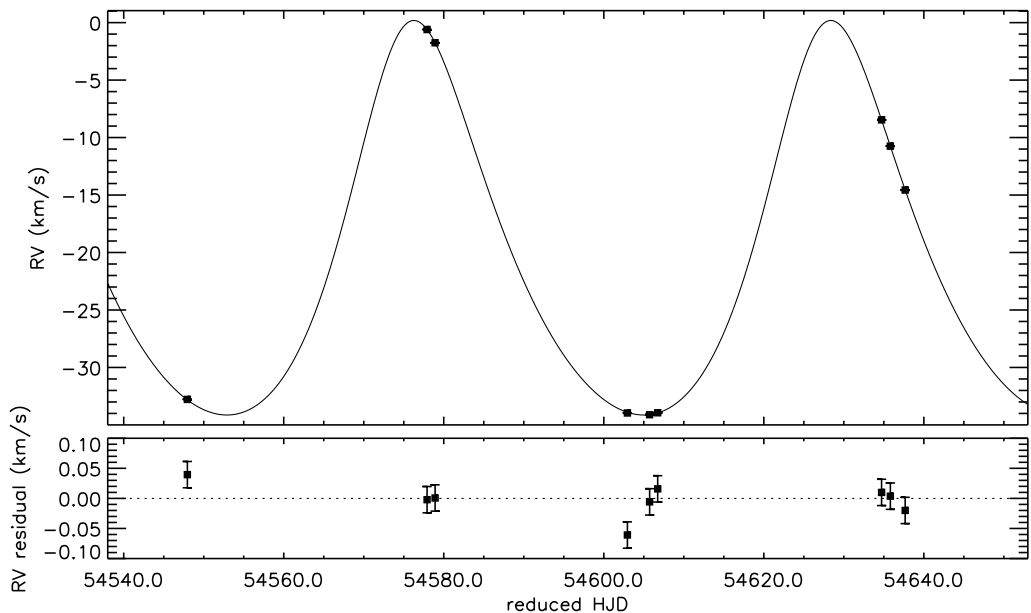}
\caption{Radial velocity measurements of the F6V star 19~Draconis. The RMS scatter of the velocity residuals is 27.5~m/s.}
\end{figure}

\subsection{$\sigma$ Geminorum}
The star $\sigma$ Geminorum (HR 2973, HD 62044, HIP 37629) has a $V$ magnitude of 4.28 and a spectral type of K1III. It is also listed as a RS CVn variable, and our measurements of spectral line widths imply a projected rotation velocity of 23.5~km/s, which may be related to its chromospheric activity level. Our RV measurements and best-fit orbit are plotted in Figure~10. As with 19~Dra, we adopt $P = 19.6044$~days from the literature \citep{mas08} and then derive: $T_{\rm peri} = 54603.597 \pm 0.189$~days, $V_0 = 44.4206 \pm 0.0502$~km/s, $K = 34.3931 \pm 0.0456$~km/s, $e = 0.0141 \pm 0.0017$, and $\omega = 5.1^\circ \pm 3.5^\circ$. The RMS scatter of the RV residuals is 25.5 m/s, as compared to a mean formal RV error bar of 29.8 m/s, and $\chi^2 = 4.50$ for $nu = 3$. The formal uncertainty on $K$ implies $\Delta K/K = 0.11$\%, but because none of our RV points are located near velocity maxima or minima, this $K$ value should be considered tentative at best and does not supersede the Massarotti results.

\begin{figure}[t]
\centering
\includegraphics[scale=0.7]{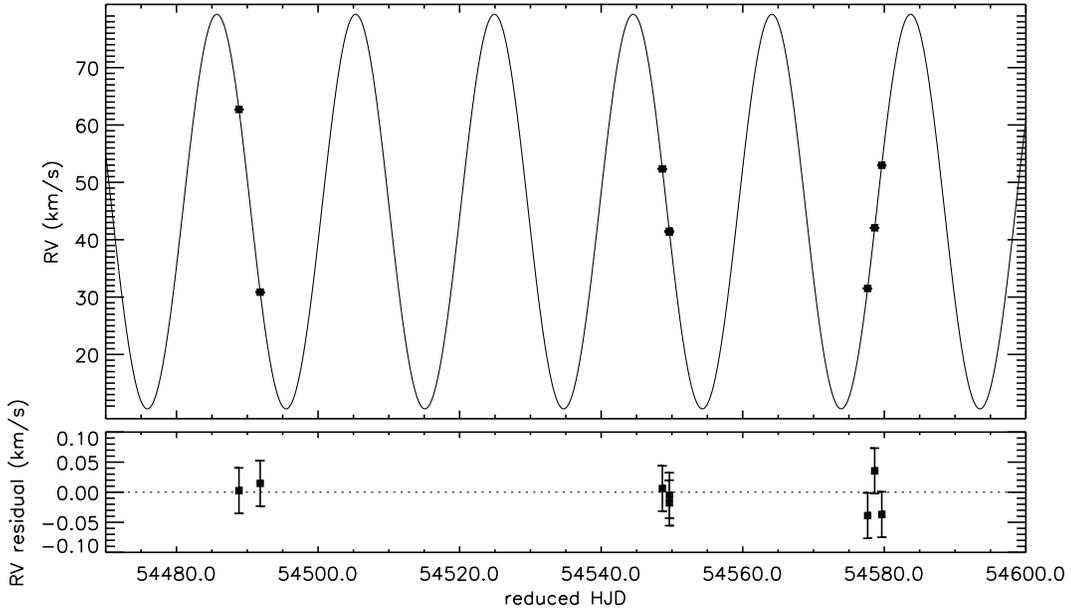}
\caption{Radial velocity measurements of the K1III star $\sigma$ Geminorum. The RMS scatter of the velocity residuals is 25.5~m/s.}
\end{figure}

Using a similar procedure as with 19~Dra, we estimate a luminosity of $23 L_\odot$ for the $\sigma$~Gem system. The spectral type of K1III implies a mass $M_1 = 4.2 M_\odot$ for the primary. The orbital inclination angle of this binary is unknown, but assuming $\sin i = 90^\circ$, we find a minimum mass $M_2 = 1.6 M_\odot$, which for a main-sequence star implies $L_2 = 5.6 L_\odot$, and thus $L_1 = 17.4 L_\odot$. The secondary is therefore at least one third the brightness of the primary, and should be visible in the composite spectrum. However, our data are not suitable for attempting this detection, as all of our measured primary velocities are near the systemic velocity, where the primary's spectral lines are likely to overlap heavily with those of the secondary.

\subsection{$\alpha$ Draconis}
$\alpha$ Draconis (HR~5291, HD~123299, HIP~68756) is an A0III star with $V = 3.65$ and an estimated $v \sin i = 26.6$~km/s. Adopting $P = 51.4167$~days \citep{els83}, we find our RV data (Figure~11) are best fit by an orbit with $T_{\rm peri} = 54527.112 \pm 0.075$~days, $V_0 = -15.5899 \pm 0.3035$~km/s, $K = 47.9340 \pm 0.2990$~km/s, $e = 0.4355 \pm 0.0042$, and $\omega = 20.8^\circ \pm 0.5^\circ$. Our determination of $\Delta K/K$ is thus 0.62\%. The mean formal error bar (172.9~m/s) and the RMS of the RV residuals (180.3~m/s) are larger than the other targets because hot stars have so few strong metal lines in their optical spectra. The orbital fit gives $\chi^2 = 9.61$ with $\nu = 5$.

From the apparent magnitude and Hipparcos parallax of this system, we estimate $L \simeq 270 L_\odot$. Because A0III is a rare spectral type (perhaps in a transitional state between two types of chemically peculiar A stars \citep{ade87}), it is challenging to determine an appropriate mass. \citet{kal04} claim $M_1 = 2.8 M_\odot$, which we will adopt. Using our mass function and assuming $i = 90^\circ$, we find $M_2 \simeq 2.6 M_\odot$, which corresponds to a A2V spectral type with $L \simeq 40 L_\odot$. The secondary could therefore account for 15\% of the total luminosity of this system.

\begin{figure}[ht]
\centering
\includegraphics[scale=0.7]{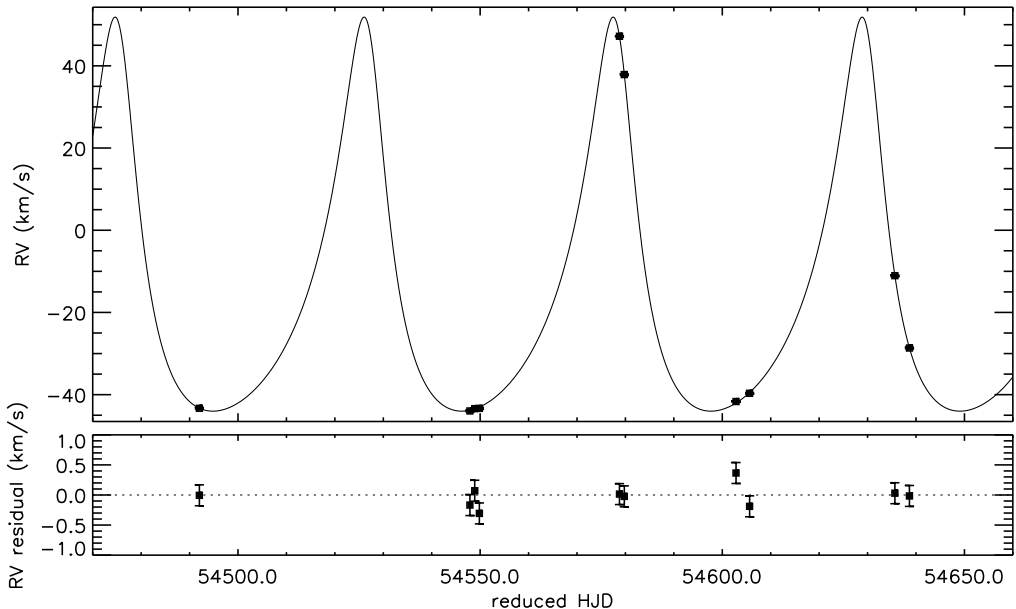}
\caption{Radial velocity measurements of the A0III star $\alpha$ Draconis. The RMS scatter of the velocity residuals is 180.3~m/s.}
\end{figure}

\subsection{$\theta$ Ursae Majoris}

$\theta$ Ursae Majoris (HR~3775, HD~82328, HIP~46853, F6IV, $V = 3.20$) is listed in SIMBAD as a spectroscopic binary, but it does not appear in the SB9 binary catalog, and \citet{wit06} found no evidence of RV periodicity above $\sim 20$~m/s. We made six observations of this star during two different observing runs. We find that its RV changed by about 180~m/s peak-to-peak, although our data do not fit any obvious Keplerian orbit (Figure~14). We hope to make further RV measurements of this star during future observing programs, to more firmly establish its binary status.

\begin{figure}[ht]
\centering
\includegraphics[scale=0.7]{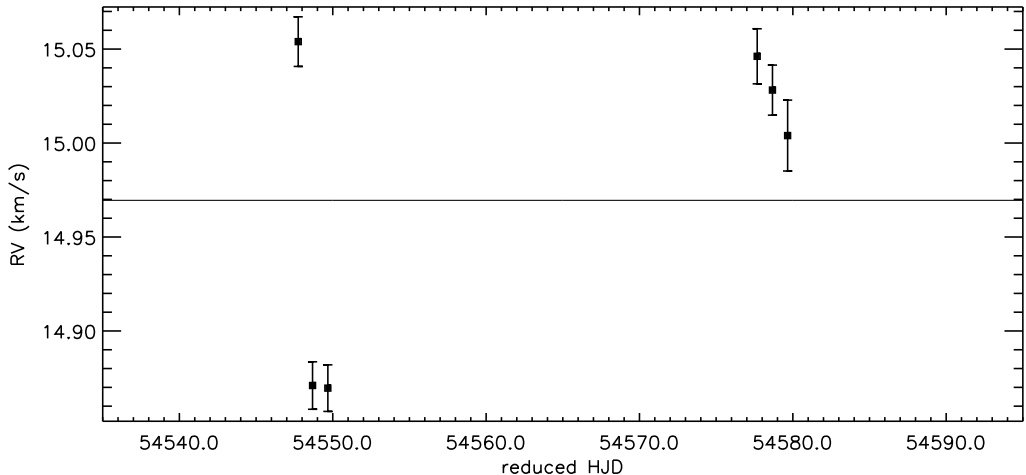}
\caption{Radial velocity measurements of the F6IV star $\theta$ Ursae Majoris.}
\end{figure}


\section{Summary}

We have described the design and operation of dFTS2, a next-generation incarnation of the dispersed Fourier Transform Spectrograph concept. This spectrometer produces $R =$~50,000 stellar optical spectra, with a wavelength calibration that enables accurate radial velocity measurement. The photon throughput and grating resolution of dFTS2 are significantly improved over its predecessor, resulting in better spectral accuracy on fainter stars, but thermal drifts inside the instrument limit its RV stability to $\sim 10$~m/s. Given a stable instrument temperature, our thorium-neon calibration data indicate that dFTS2 can reach RV stability better than 1.5~m/s RMS. These temperature issues will addressed in a future dFTS instrument, at which point the RV accuracy will be limited primarily by photon statistics and intrinsic stellar velocity variability.

We have shown that dFTS2 can accurately measure the RV curves of single-lined spectroscopic binaries, with formal errors on the velocity amplitude $K$ of approximately 0.1\% in some cases. Table 1 summarizes the derived orbital parameters. However, some of our results must be considered preliminary because of the relatively small number of data points. When the third-generation dFTS has been constructed and deployed, we anticipate being able to significantly improve the RV measurements of spectroscopic binaries, and in conjunction with astrometry from spatial interferometers, determine the masses of stars with unprecedented accuracy.

\begin{center}
\rule{150pt}{0.5pt}
\end{center}

\begin{deluxetable}{lllllll}
\tabletypesize{\scriptsize}
\tablecaption{The SB1 orbital parameters derived from dFTS2 observations.}

\tablehead{\colhead{star} & \colhead{$P$ (days)} & \colhead{$T_{\rm peri}$ (reduced JD)} & \colhead{$V_0$ (km/s)} & \colhead{$K$ (km/s)} & \colhead{$e$} & \colhead{$\omega$ (${}^\circ$)}}
\startdata
19~Dra		&52.1089			&$54522.258 \pm 0.121$	&$-20.5299 \pm 0.0649$		&$17.1590 \pm 0.0344$		&$0.2218 \pm 0.0020$		&$339.1 \pm 0.6$	\\
$\sigma$~Gem		&19.6044		&$54603.597 \pm 0.189$		&\phm{$-$}$44.4206 \pm 0.0502$		&$34.3931 \pm 0.0456$		&$0.0141 \pm 0.0017$		&\phn\phn $5.1 \pm 3.5$ \\
$\alpha$~Dra		&51.4167		&$54527.112 \pm 0.075$		&$-15.5899 \pm 0.3035$		&$47.9340 \pm 0.2990$		&$0.4355 \pm 0.0042$		&\phn $20.8 \pm 0.5$	\\
\enddata
\end{deluxetable}

\pagebreak

\acknowledgments

We are grateful to the day crew at Steward Observatory --- Jeff Fearnow, Dave Harvey, Bob Peterson, Gary Rosenbaum, and Bill Wood --- for their assistance with the transport and installation of dFTS2, and we thank telescope operators Geno Bechetti, Dennis Means, and Peter Milne for their expertise in operating the telescope on our behalf. We also express our appreciation to the Director of the Steward Observatory for granting us telescope time over an extended period.

We are greatly indebted to the skilled instrument builders in the USNO Machine Shop --- Gary Wieder, Dave Smith, Tie Siemers, and John Evans --- for fabricating all of the custom optomechanical elements of dFTS2, as well as the thermal enclosure. We also thank the USNO Astrometry Department for travel support and salary support during the initial stages of this observing program, and thanks also go to the USNO Time Services Division for lending us packing crates for shipment of our instrument to Kitt Peak. 

This research has made use of the SIMBAD database, operated at CDS, Strasbourg, France; NASA's Astrophysics Data System; and the SB9 catalog of \citet{pou04}. Richard O. Gray is to be commended for making his SPECTRUM codes so easy to install and use. 


\end{document}